# Should Cyberspace Chat rooms be closed to protect children?

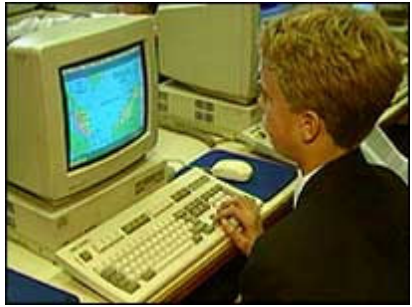

Source:http://news.bbc.co.uk

**Introduction**

The explosion of people networking in cyberspace, disseminating terabytes of information is being promoted through the use of broadband, bluetooth technology, and wireless mobile computing facilities such as laptops, PDA's and now smart-phones. New communities within such venues as virtual chat rooms, Forums, e-classrooms videoconferences, discussion groups, newsgroups etc. are being created daily and even hourly.

This is raising issues of cyberethics concerning privacy, security, crime, human needs, e-business, e-healthcare, e-government and intellectual property among others that need to be evaluated and reflected upon.

With this new freedom come new moral and ethical responsibilities, which raise questions as to whether anything can be published or whether there should be restrictions. This paper addresses one specific area, that has come into the public eye recently, the closure by Microsoft of all of its free[1] chat rooms.

Microsoft claim with some evidence, that certain users are misusing chat rooms to contact children, and lure them into dangerous relationships. Microsoft say that it is impossible for them to make their chat rooms 100% safe as they are not able to moderate every single message that appears. Therefore they claim that the only safe action is to close down all of their free chat rooms so as not to put children at risk.

Four arguments in favour of closure and six arguments against closure have been examined in some detail (see Appendix 1). Various methods of ethical analysis (Appendix 2 and 3), and relevant lessons about present laws, the future, policy vacuums and conceptual issues are addressed.

---

[1] MSN has announced that it is closing its chatrooms throughout Europe, Latin America, the Middle East and most of Asia. Similarly, in the United States, unmoderated chatrooms will only be available with a credit card subscription.
**http://www.maxpc.co.uk/default.asp**

**Arguments for and against chat room closure.**
The four arguments in favour of closing chat rooms are that there is no other option, that it is the only practical solution, that the misuse of the minority must be dealt with whatever the cost, and finally that it is the only responsible course of action.

These arguments are all shown to be invalid for a variety of reasons (see Appendix 1). The fact that chat rooms are open to abuse does not mean they should of necessity be closed (The slippery slope fallacy[2]). The fact that moderated chat rooms cannot always be made available does not mean unmoderated ones should close (the "next best" fallacy). The fact that innocent activities can be abused does not mean that all innocent activities should be stopped (the 'minority misuse' fallacy). Nor is it established that closure is the only responsible action (the 'begging the question' fallacy[3]).

Similarly the six arguments against closure have been analysed in some detail, and these involve the dangers of forcing children into less regulated areas, that closing chat rooms may prompt children to give out their personal contact details, that it is pointless as people will simply move to other chat rooms, that the responsibility for protecting children does not lie with Microsoft but with their parents, that it is an over reaction to a 'miniscule risk'[4], and finally that the cause of the problem has been mis-identified.

It is shown that four of these arguments are invalid and two valid (Appendix 1). It is not established that closing chat rooms will force children into other more dangerous areas (the 'for fear of finding something worse'[5] argument). Neither is it established that closing chat rooms will of necessity force children to disclose contact details to maintain relationships (the 'forced disclosure' argument). It is not valid to maintain that chat rooms should not be closed because people can move to different chat rooms. This depends on the premise that not all chat rooms can be closed. The last of the four invalid arguments is that the risk is so small it should be ignored. This is an example of the 'begging the question' fallacy. This is not say that there is no danger, just that the arguments are not logically conclusive.

The first of the two valid arguments is that it is the parents' responsibility to monitor children in chat rooms, and that in closing chat rooms Microsoft have usurped this role which really belongs to the parents.[6] The second is the 'confused cause' argument. Chat rooms should not be closed because chat rooms are not the problem. The problem is people who misuse chat rooms. The solution should be to deal with the individuals and so deal with the correct cause.

---

[2] Herman T. Tavani, Ethics Technology, P 70
[3] Herman T. Tavani, Ethics Technology, P 71

[4] Alison Perret, **http://www.spiked-online.com/sections/technology/index.htm**, 3rd October 2003
[5] Hilaire Belloc, Cautionary Verses, G. Duckworth & Co. Ltd, 1939
[6] Parent's action of monitoring their children in chat rooms may be argued as a breach of their children's right to privacy.

To summarise it has been shown that all the arguments examined for the closure of chat rooms have been shown to be invalid. Two of the six examined arguments for non-closure have been accepted as valid.

**Ethical theories and traditional solutions.**
Consideration is now given to the position of chat room closure in the context of ethical theories and traditional solutions. The details of this are contained in Appendix 2.

Performing an Aristotelian analysis which is based upon the "virtue and the vices" of the situation, it is discovered that adult responsibility, childhood innocence and parental responsibility are weighed against the vices of child abuse and the possible self serving protectionism of Microsoft. Weighing factors like this against each other is inevitably a subjective process from which no valid conclusion can be easily drawn.

The utilitarian analysis of Bentham weighs the benefits of child protection and the prevention of exploitation against the loss of rights of association, expression and dissemination and the possible loss of revenue by Microsoft. Bentham's principle of the greatest happiness for the greatest number suggests that the rights of the majority counteract the misuse of the minority. This indicates that chat rooms should not be closed.

Kant's categorical imperative places the actions of the individual as only being valid if they can be translated to the actions of society as a whole. If a rule can apply to everyone then it can be applied to an individual. The conclusion is that the actions of abusers should be restricted and the genuine use of the majority allowed. One should not be traded off against the other.

Traditional solutions also point to the non-closure of chat rooms, as there are many instances where it has been established that facilities/methods etc. should not be stopped merely because they can be abused. Examples of this have been cited in Appendix 2.

On balance, applying ethical theory analysis it is concluded that chat rooms should not be closed on the basis of protecting children.

**Further Analysis Techniques**

Full details of the following analyses are found in Appendix 3.

A **roles and responsibility** analysis has shown that there are five roles involved, Adult user, Mis-user, Child user, Parent and Microsoft. Responsibilities to protecting children and others from abuse are balanced against freedom of expression, dissemination and association. This analysis provides no clear conclusion.

The **stakeholder** analysis identifies the same groups as stakeholders and seeks to balance the benefits of closure for each against the harms. Again no clear-cut conclusion could be drawn from this.

The **professional standards** analysis shows that the closing of chat rooms amounts to denial of accessibility, freedom of expression and dissemination and association, which is a contravention of many standards. On the other hand standards also require that there should be no harm to other people and the respect for fellow human beings.

However, it is argued by many that codes of conduct and professional ethics are not sufficient to determine or regulate ethical actions on the Internet. This is because they are not compulsory and therefore have no bite, have loopholes, are vague, are not enforceable over the World Wide Web and are inconsistent[7]. It is not clear how to enforce such codes with respect to individuals. In view of present and future dangers it might be argued that computer professionals and the Internet or rather the ISP's need to be subjected to some form of regulating body (the same as the BMA and the Law society) plus additional legislation that will limit these dangers before they get even further out of hand.

A **justice, respect and equality** analysis also resulted in a no clear-cut answer. Respecting the innocence of children is difficult to balance against the respect for freedom of expression and the injustice caused by closing the chat rooms. The rights of the majority are compromised if chat rooms are closed.
All of these analyses are inconclusive because the balancing of rights is often a subjective process with no unquestionable solution.

**Present laws and Policy Vacuums**

The Regulatory Investigative Powers (RIP) Act already allows information held by ISPs to be accessed by the Police to track cyber movement in this country. Echelon is a programme in use by the authorities that intercepts and reads billions of emails everyday. Police regularly use entrapment techniques to catch criminals on the Internet. It has been argued that this is breaching privacy laws by law enforcement agencies and there are those who argue that the cure is worse than the illness. Are those who enforce the law above it? The answer must be no.

---

[7] Dr.M.Wald,INFO3001,Course Notes week 4

As a consequence, many today regularly use anonymity tools such as Pretty Good Privacy, a highly sophisticated encryption technology, to protect themselves from Internet eavesdropping. Software is commonly used to erase Internet trails from PCs to prevent tracking and preserve privacy. All of this makes it more difficult to bring Internet abusers to justice, in particular if they reside in countries, which take an ambivalent view to such crimes.

On the issue of chat room closure the most obvious question to ask is who is responsible if a malevolent person infiltrates the chat rooms and harms children. Is the chat room provider responsible in any respect? This is yet to be tested in a court of law. Moving down one level, does responsibility also fall upon those who have written the program that allows chat rooms to run? Is the software developer responsible for the misuse of his creation? Delving down further one can ask whether those who provide the hardware on which this runs; the routers and servers are also responsible for allowing their equipment to be misused in this way. Are ISPs responsible for hosting this? These are areas that still need to be tested.

**Future considerations**
The issues can only be expected to become more sharply focussed with the advent of new technologies. Progress cannot be allowed to be halted because of malevolent influences. Already some chat rooms make use of avatars, which are small cartoon like characters chosen by the user to express their projected personality and feelings. The next generation software can be expected to expand on the visual side and it can be anticipated that within 10 or 20 years a fully developed virtual environment may accompany chat with all of the potential dangers that would bring. The potential for disguise and misrepresentation may be secondary to greater visual psychological damage and a disturbance of the psychological nature of the child. It will therefore be necessary to adopt some solution, which can circumvent the pernicious dangers before they become exploited. Maybe some restrictions as to adopted personality or avatar could be introduced with access to only appropriate persons.

One can foresee a time when artificial intelligence has so advanced that it becomes impossible to distinguish between a software chat room partner and a human chat room partner. In this event Microsoft may be able to provide the perfect environment for children to chat to friends and be completely protected as they build relationships with harmless chatbots.[8]

However this raises further issues which concern responsibility should the 'chatbots' prove unreliable or malicious. Who carries the ultimate responsibility for a robot, since the robot has no responsibility of its own? It may be necessary to have constant vigilance within chat rooms assigned to other software. One can envisage a 'watcherbot' given the task of policing the chat room environment. Should the 'watcherbot' go wrong one is left with the question who watches the 'watcherbot'? Infinite regress is the result of this way of thinking.

---

[8] Some may see the use of chatbots as a deception played upon the child if the child is not aware of whom or what it is talking to. One way around this would be to require all chatbots to declare themselves as not 'real'.

One can foresee a danger in the future of malevolent software being written which is designed to snoop into chat rooms and discover user types, typically children, and to set up relationships with them and perform the grooming process automatically. The difficulty here is proving a malevolent intent from a software 'chatbot'. The intention lies with its originator rather than with the software.

The far distant future may well see a very different view of the Internet from what we have in nascent form today. The increasing sophistication of 'webbots' and 'chatbots' may so take on their own personalities and characters and develop and show insight, imagination, creativity and self existence that it becomes impossible by any means to distinguish them from a living sentient being. When software has reached this stage the Turing test will have been well and truly passed, and the question will have to be answered as to whether there is such a thing as self-existent 'weblife'. The issue for those in the distant future will be what rights to accord this kind of 'weblife'. Could a malevolent 'weblife' be responsible for its own actions and would it be just to terminate it if it committed a crime. We leave this as an exercise for the reader.

**Conclusion**
The question of freedom on the Internet and the particular issue of chat room closure to protect children have been examined. The arguments from Microsoft in favour of closure were shown to be lacking the rigour, which an inescapable conclusion would demand. Microsoft's arguments are also tainted by accusations that they are protecting their own interests more than the interests of others, and further criticisms pointed towards the usurping of parental rights in the area of policing chat rooms. The curtailing of the rights of freedom of expression, dissemination and association of the majority cannot be overran for the sake of a real but relatively low risk danger. After all, many children are knocked down and killed by motorcars every day of the year but this is not a valid reason to ban motorcars.

Whatever the rights and wrongs of chat room closure are, all are agreed that children need to be protected. How this is best achieved is the question. Should this be left to the parents or to Microsoft?

Chat rooms are global communication platforms, and it is at present difficult for global rules to be enforced effectively. What may be acceptable practices in one culture could give offence to another. Even if we had computer ethics acceptable to all cultures, how can they be enforced? These issues need to be examined in depth on a global basis if universal laws and ethical guidelines in cyberspace are to be established.

One area of cyberethics has been examined and analysed. The conclusions are not simplistic nor easy and different approaches have yielded different answers. The cyberspace explosion has thrown up myriads of such ethical problems, and there are no clear-cut answers to satisfy everyone. Such cyberethic issues will continue to challenge us well into the future.

**Bibliography:**
1. Matthew Whittingham, Head of Customer Satisfaction, MSN UK
2. Camille de Stempel, Director of Policy, AOL UK
3. **www.maxpc.co.uk**
4. Alison Perrett, **http://www.spiked-online.com/sections/technology/index.htm**
5. John Carr Internet Adviser from children's charity, NCH
6. Alex Kovach, Managing director, Lycos UK & Ireland
7. Hilaire Belloc, Cautionary Verses, G. Duckworth & Co. Ltd, 1939
8. Rachel O'Connell, Director of Research, Cyberspace Research Unit, University of Central Lancashire
9. Parul Amlani, Sunday October 12, 2003, The Observer
10. Terrel Ward Bynum et al, Computer Ethics and Professional Responsibility, Blackwell Publishing, 2004
11. Herman T. Tavani, Ethics & Technology, Wiley, International Edition, 2004
12. Ben Schneiderman, Leonardo's Laptop, MIT Press, 2003
13. **www.acm.org**
14. **www.computer.org**
15. **http://www.ieee.org**
16. British Computer Society
17. Dr. M. Wald, Info 3001 Course Notes
18. **Micah L. Sifry,** Tripping on Internet Populism, http://www.thenation.com , 1st March 2004
19. http://www.oreillynet.com
20. Paul Harris, The Observer, 2nd February 2003
21. The Guardian, Leader, 11th June 2002
22. **http://www.cyber-rights.org/reports**
23. Simon Davis, Director of Privacy International
24. **http://news.bbc.co.uk/1/hi/magazine/3135620.stm**

# APPENDIX 1

**ARGUMENTS IN FAVOUR OF CLOSURE**
Analyzing the arguments in favour of closing chat rooms they fall into four categories.

**1 The 'we have no other option' argument**
*"It is impossible to monitor every single message from millions of users, and moderation doesn't prevent spammers from sending pornographic spam messages to chat rooms".*
**Matthew Whittingham**
**Head of Customer Satisfaction, MSN UK**

**Analysis of Argument 1**
Premise 1     Paedophiles stalk chat rooms
Premise 2     Microsoft cannot make chat rooms 100% safe
Conclusion    Chat rooms should be closed down

This is Microsoft's own argument. "There is no other option since it is impossible to monitor every single message."

This is an example of the slippery slope argument. Microsoft has assumed that the use of chat rooms inevitably leads to a child abusive slippery slope. However although it may be true that chat rooms are open to abuse it does not necessarily follow that chat rooms must be closed. There may be other solutions, which may deal with that abuse rather than simple closure. In addition it might be argued that premise 2 has not been fully demonstrated and that principles of good practice might be discussed across the industry may be in place which could make chat rooms 100% safe.

Therefore the argument is fallacious on two grounds; firstly that the strength of the syllogism is invalid, and secondly premise 2 has not been established.

**2. "The Second best" argument**
*"MSN's decision to close its open chat rooms has been criticized because it is feared that children will simply go to other unmoderated chat rooms. Moderated chat rooms are clearly more appropriate for kids, but if you cannot offer moderated ones, then shutting them down is a good idea"*.
**Camille de Stempel**
**Director of Policy, AOL UK**

**Analysis of Argument 2**
Premise 1: Moderated chat rooms are best for kids,
Premise 2: Can't always offer moderated chat rooms,
Conclusion: Shutting them down is second best

It has not been unimpeachably shown that either of these premises are logically or factually valid. Furthermore, it is not a logical conclusion to identify closure as the next best option. It may be that there are other options that are better but these have not been discussed.

**3 The Argument of Minority Misuse**
"*The stated aim of the move is to protect users from spam and safeguard children from inappropriate communications. In a statement Gillian Kent, the Director of MSN UK, said: "As a responsible leader we felt it necessary to make these changes because online chat services are increasingly being misused."*
**www.maxpc.co.uk**

**Analysis of Argument 3**
Premise 1: Children may be engaged in online chat
Premise 2: Online chat services are increasingly being misused by a minority
Conclusion: Online chat should be closed

This argument is similar to one advanced in an article by Allison Perrett, "Imagining pornographers everywhere". She relates that her daughter was engaged in the innocent activity of bathing in a public fountain. She further relates how a member of staff told her actions were wrong because a perverted photographer might put innocent photographs on the wrong web site.

This is what is meant by the minority misuse argument. In the context of this essay the innocent activity is the use of a chat room, which may be perverted by a pornographer. The conclusion is that the innocent activity must be restricted.

This argument is fallacious because it proves too much. If allowed it restricts any and every innocent activity on the grounds that some perverted minority might misuse it.

**4 The argument of responsible action**
*"MSN took the only responsible option open to them, which was to get out of the market altogether".*
**John Carr Internet Adviser from children's charity, NCH**

**Analysing Argument 4**
Premise 1      Running chat rooms is becoming irresponsible
Premise 2      Irresponsible activities should be stopped (implicit)
Conclusion     It is responsible to close chat rooms

Although the construction of the argument is valid it is not clear that all the premises are sound. The premise that running chat rooms is becoming irresponsible is begging the question. The argument has circular reasoning and instead of establishing that running chat rooms is irresponsible the author assumes this to be the case.

**Summary or arguments for closure**

|            | Name                  | Type                | Argument | Premises |
|------------|-----------------------|---------------------|----------|----------|
| Argument 1 | We have no other choice | Slippery Slope    | Invalid  | Invalid  |
| Argument 2 | Second Best Action    | Begging the question | Invalid  | Invalid  |
| Argument 3 | Minority Misuse       | Proves too much     | Invalid  | Valid    |
| Argument 4 | Minority Misuse       | Proves too much     | Invalid  | Valid    |
| Argument 5 | Responsible Action    | Begging the question | Valid   | Invalid  |

It will be seen from the summary table that no argument has both valid premises and valid construction. In this case all arguments fall and we have to conclude that there are in fact no valid arguments for the closure of chat rooms.

**ARGUMENTS AGAINST CLOSURE**

Opposing this view are many of Microsoft's competitors whose chat rooms are still open. They claim that Microsoft's real reason is to protect themselves against legal claims. We have discovered six arguments against closure of chat rooms, which will be examined in turn.

**1 The " For Fear Of Finding Something Worse"[9] argument**
*By switching off chat rooms Microsoft look like they're taking the moral high ground but in reality this is an irresponsible move. This will only drive chat rooms underground, and therefore users and more specifically children, into more dangerous environments.*
**Alex Kovach**
**Managing director, Lycos UK & Ireland**

**Analysis of Argument 1**
| Premise 1 | Closing down chat rooms force children to go other areas |
|---|---|
| Premise 2 | Other areas are more dangerous |
| Conclusion | Closing chat rooms is more dangerous for children |

The premises of this argument are not logically and inevitably sound. Closing down a chat room may or may not force children into other areas. Furthermore other areas may be more dangerous but not all areas. The wording of premise two is ambiguous as to which area is dangerous. The validity of this argument is sound in so far as the conclusion forms a valid syllogism, and cannot be contradicted on this basis. However the fallacy of ambiguity is also present in this argument because the term "other areas" in premise 1 may not mean the same as "other areas" in premise 2. There is therefore an inherent weakness in this argument.

**2 The "forced disclosure" argument**
*We believe that in the short term this move could spark unease amongst children who use MSN chat regularly and may prompt them to be more willing to give out personal contact details, including phone numbers.*
**Rachel O'Connell**
**Director of Research, Cyberspace Research Unit, University of Central Lancashire**

**Analysis of Argument 2**
| Premise 1 | People build relationships in chat rooms |
|---|---|
| Premise 2 | Relationships will be maintained when chat rooms close |
| Conclusion | People will disclose contact details to maintain relationships |

An examination of the premises leads us to conclude that while premise one is true in fact, premise two is neither necessarily nor logically true. The argument is that closing chat rooms will lead to increasing danger as people are forced to disclose contact details which they would normally keep secret rather than lose the relationships that they have build up. This is a powerful argument from a practical standpoint but does not have structural or logical validity. It may well be a keen insight into the understanding of human nature, but is not logically forced upon us.

**3 The 'you can't stop them talking' argument**
*Anyone who seriously believes that MSN's chat room closures are a positive step to improve Internet safety is basking in their own naivety. For one thing, it just is not feasible to close all chat rooms given that there are at least several hundred thousand of them worldwide. It follows*

---
[9] With apologies to Hilaire Belloc

*therefore that chat rooms have such a widespread following that any closures will redirect users to an alternative service.*
**Parul Amlani**
**Sunday October 12, 2003**
**The Observer**

**Analysis of Argument 3**
| | |
|---|---|
| Premise 1 | There are many chat rooms that people can use |
| Premise 2 | You can close some but not all chat rooms |
| Conclusion | People will move to different chat rooms |

This is based on the premise that it is impossible to close all chat rooms. This is a practical impossibility rather than a theoretical or logical impossibility. Even so it is not necessarily true as it may be possible to ban all chat rooms sometime in the future. If this happens then the argument will fail. In addition it is quite likely that people will move to different chat rooms but again this is not a logical necessity, in which case the argument itself is invalid.

**4 The argument of misplaced responsibility**
*It stands to reason that the biggest obligation and responsibility in preserving child safety rests firmly with the parents..... Society is not at the mercy of chat room malice. There is a multitude of things that can be done by anyone who chooses to assume responsibility.*
**Parul Amlani**
**Sunday October 12, 2003**
**The Observer**

**Analysis of Argument 4**
| | |
|---|---|
| Premise 1 | Children should be monitored in chat rooms |
| Premise 2 | Parents are responsible for children |
| Conclusion | Parents are responsible for monitoring their children in chat rooms |

An examination of the premises shows them to be self-evident and incontrovertible. Likewise the conclusion is validly drawn from the premises. The claim is that by closing its chat rooms Microsoft has usurped the authority and responsibility of the parents for monitoring their children. It is not Microsoft's role to be that of parent.

However it is not entirely clear that this absolves Microsoft of all responsibility. Although they are not responsible for children they are responsible for their chat rooms and have the right to close them if they so wish – especially if they think they might be sued. But closing the chat rooms for the sole purpose of protecting children may not validly be their role. Somebody else is charged with that responsibility.

**5 The overreaction to miniscule risk argument**
*This will be one more nail in the coffin for a truly free and open internet...... The internet is on course to becoming a shadow of its former self, due to a collective lack of trust in ourselves and in those around us...... Microsoft is purporting to take the moral initiative, by removing the burden of responsibility from the hands of concerned parents. But its chat room closures are a sad overreaction to a limited problem that could be checked by greater parental supervision and awareness of our children's computer activities.*
**Alison Perrett works as a copyright assistant at the Tate gallery and has an MSc in Electronic Publishing from City University. 3 October 2003 http://www.spiked-online.com/sections/technology/index.htm**

Allison Perrett does not believe that Microsoft can take the moral high ground by taking responsibility out of the hands of parents. She argues that it is more a matter of smart marketing on the part of Microsoft.

**Analysis of Argument 5**

| | |
|---|---|
| Premise 1 | Pornographers use chat rooms |
| Premise 2 | Ordinary people vastly outnumber pornographers |
| Conclusion | The chances of meeting a pornographer are so miniscule as to be almost non-existent |

Allison Perrett's argument is that chat room closures are a "sad overreaction to a limited problem" An analysis of the individual premises shows them both to be factual and correct. However the conclusion here that "the chances are so miniscule." is at least overstating the case and at best a debatable conclusion because one man's miniscule risk is another man's serious risk. The argument hinges upon the almost emotive term miniscule to mean something, which can be ignored. This is begging the question.

**6 The misplaced cause argument**
*" The police could work more with chat service providers and phone companies to secure more convictions and prosecutions"*
**Parul Amlani**
**The Observer**
**Sunday October 12, 2003**

**Analysis of Argument 6**

| | |
|---|---|
| Premise 1 | Chat rooms are not dangerous by themselves |
| Premise 2 | The danger comes from those who use the chat rooms |
| Conclusion | Dealing with the people who use the chat rooms is the answer |

The argument here is that we need to distinguish the correct cause and that Microsoft has failed in its arguments because it has used the false cause fallacy. Microsoft identifies chat rooms as the problem whereas in reality the opponents say that the problem is with some people who use the chat rooms. The conclusion that Microsoft come to that they should close the chat rooms is only correct if the chat rooms are the problem. However if that is not the case then the solution is not to close the chat rooms but to deal with the individuals.

**Summary of Arguments Against Closure**

| | Name | Type | Argument | Premises |
|---|---|---|---|---|
| Argument 1 | For fear of finding something worse | | Valid | Invalid |
| Argument 2 | Forced Disclosure | | Invalid | Invalid |
| Argument 3 | You can't stop them talking | | Invalid | Invalid |
| Argument 4 | Misplaced Responsibility | | Valid | Valid |
| Argument 5 | Overreaction to miniscule risk | | Debatable | Valid |
| Argument 6 | The Misplaced Cause | | Valid | Valid |

An examination of the table above shows that most arguments do not have valid premises and valid constructions. However there are two arguments, number 4 and number 6 which have both valid premises and constructions. These arguments present us with a clear conclusion that chat rooms should not be closed on the basis of protecting children.

## APPENDIX 2

**Applying Ethical Theory Analysis**
A number of principles have been proposed and grown to be accepted over many years. The earliest guide in ethical analysis was Aristotle.

**Aristotelianism**
Aristotle's values were based upon an analysis of vice and virtue. In the case of chat rooms this can be summarised in the following table where five users have been identified.

| Aristotelian Analysis: Close chat rooms to protect children | | |
|---|---|---|
| **Participant** | **Virtue** | **Vice** |
| Adult User | Responsibility | NA |
| Mis-user | NA | Child Abuse |
| Child User | Innocence | NA |
| Parent | Responsibility to Children | NA |
| Microsoft | Child Protection | Self-serving protection |

It is difficult to weight one virtue against another vice or to give them weighting factors. Any such attempt is bound to suffer from the objection that it is subjective. There is no compelling answer from this analysis.

**Utilitarianism**
Bentham's utilitarianism, which proclaims the greatest happiness for the greatest number can be used to analyse the Chat room case. As Bentham put the happiness of the majority over the minority, it should be clear that utilitarianism would never allow the happiness of the majority to be curtailed by the malevolence of the minority.

The majority of chat room users are acknowledged to be decent upright folk and removing their chat room rights for the sake of combating a minority activity would be considered inappropriate.

| Utilitarian Analysis: Close chat rooms to protect children | | |
|---|---|---|
| **Participant** | **Benefit** | **Harm** |
| Adult User | NA | Loses rights of association and expressions and dissemination |
| Mis-user | NA | Loses rights of association and expressions and dissemination |
| Child User | Protection | Loses rights of association and expressions and dissemination |
| Parent | Child protection | NA |
| Microsoft | Child Protection/ Self Protection | Loss of revenue |

**Kantian View**
Kant made the point the every moral action can be tested by generalising it to the whole of society. If it stands that test then it can be applied in the particular case. If we were to ask is it right to lie, Kant would require us to consider the implications of everyone lying and if we concluded that this would cause a dysfunction in society then no one person can lie. As it cannot work in the general case it should not be allowed in the specific case. The stopping of innocent activities cannot be a moral imperative. Rather the restrictions need to be placed upon the misuse of the chat room by the perverted user. The conclusion is that it would be right to stop chat room misuse, but not right to stop chat rooms.

**Traditional solutions and analogies**
The argument that if something is open to misuse it should be closed does not have a large following in practice. Camera use can be abused by photographing inappropriately. Videos or CDs can be abused by infringing peoples copyright etc. Many non-prescription drugs such as paracetamol can be misused. But this is not considered a valid argument for banning any of these things.

Alternatively there have been instances where services have been closed through misuse. The case of premium telephone lines which have tricked people into running up large bills have been closed by BT. However it is not clear cut that this is a valid precedent. For one thing a premium telephone number has no alternative legitimate use as chat rooms do. In this way we are not infringing the rights of a lawful majority.

It is concluded that traditional solutions point towards the non-closure of chat rooms.

# APPENDIX 3

**Further Systematic Analysis techniques**
There are various techniques we can use to apply to the ethical question of chat room closure, this include an analysis of roles and responsibilities, stakeholder analysis, professional standards and ethical theory. To perform these we need to identify the user groups.

- First there were those who use chat rooms for good and useful purposes such as building communities, building relationships, sharing data etc.
- Secondly there were children who use it for the same purposes.
- Thirdly there is a minority who misuse and abuse the services and other members of the chat rooms.
- Fourth we might also consider the parents of the children.
- Fifth there is the chat room host, in this case Microsoft.

The largest group is the first who have rights of freedom of expression association and dissemination. These would be compromised by closing chat rooms.
The second group is in the same situation, and the third group are those whose activities we seek to limit/ remove. The roles and responsibility analysis is set out in the table below.

**Roles and responsibilities analysis**

| Role | Responsibilities | Rights | Rights Respected? |
|---|---|---|---|
| Adult User | To protect themselves and others from misuse or abuse | Freedom of expression, dissemination and association | Not if the chat room is closed |
| Mis-user | To protect themselves and others from misuse or abuse | Freedom of expression, dissemination and association | Not if the chat room is closed |
| Child User | NA | Freedom of expression, dissemination and association | Not if the chat room is closed |
| Parent | Protect Children | NA | NA |
| Microsoft | Protect the chat rooms | Rights of ownership | YES |

This analysis suggests conflicting rights and responsibilities but no clear conclusion.

**Stakeholder analysis**
Stakeholders are those who are significantly benefited or harmed by a particular course of action as shown below.

| Issue: Chat rooms closed to protect Children | | |
|---|---|---|
| **Stakeholder** | **Benefits** | **Harms** |
| Adult User | None | Curtailment of freedom of expression, dissemination and association |
| Mis-user | None | Curtailment of freedom of expression, dissemination and association |
| Child User | Protection against misuser/'abuser | Curtailment of freedom of expression, dissemination and association |
| Parent | Protection of Children | Possible usurping of parental right |
| Microsoft | Protection against legal claims | Bad publicity<br>Loss of users and advertising revenue |

Again competing benefits and harms are difficult to weigh against each other and although it might be possible, which falls on the heavier side, there is no clear cut compelling conclusion from this approach.

**Professional standards and ethical codes analysis**
Microsoft's actions can be analysed against Professional codes of ethics and professional conduct. Although it is not known whether Microsoft is a member of any professional body with a code of conduct, nevertheless most statements of professional conduct encompass diverse activities. Some have identified four main areas of ethical code[10]. These concern privacy, accuracy, property and accessibility. However this is too narrow a focus as ethical code and conduct ranges far beyond those four issues. The issue at hand, that of closing chat rooms, concerns what amounts to a denial of accessibility, expression, dissemination and association and therefore would not accord with many of the prime ethical issues.

Three of the "Ten Commandments of Computer Ethics" from the Computer Ethics Institute deal with harm to other people, social consequences and a respect for fellow humans. It could be argued that these latter concerns in the case of the dangers posed in chat rooms outweigh the considerations of accessibility expression, dissemination and association. It appears that there is no clear conclusion from an analysis of ethical codes.

The code of conduct of the British Computer Society directs that there should be no discrimination that all should be treated with dignity and respect, and the European convention on Human Rights should apply where relevant. In particular section 5

*" You are encouraged to promote equal access to the benefits of IS by all groups in society, and to avoid and reduce ' social exclusion' from IS wherever opportunities arise." (BCS Code of Conduct section 5)*

It is not clear how the closure of chat rooms can be held consistent with promoting access and reducing social exclusion.

**The ethical point of view today**
The ethical point of view is how do these proposals measure us in the areas of Justice, respect and equality. We take as axiomatic that all human beings are equal in the eyes of justice. This requires us to respect each person's needs and rights. In the case of community chat rooms a large number of people are involved.

|                          | **Justice**                        | **Respect**                                                                      | **Equality**                                                                  |
| ------------------------ | ---------------------------------- | -------------------------------------------------------------------------------- | ----------------------------------------------------------------------------- |
| **Closing Chat rooms**   | Unjust to innocent chat room users | Respecting the innocence of children                                             | Does not distinguish between the innocent and the guilty. Affects all.        |
| **Keeping Chat rooms open** | No Injustice involved           | Respects the freedom of expression, association and dissemination of Chat room users | Open for all.                                                                 |

---

[10] (Mason 1986, "Four ethical issues of the information age" – Terrell Ward Bynum et al P 142)

By closing the chat rooms the rights of the majority would be compromised. On the other hand the possible abuse of children is a severe curtailment of their human rights. There is therefore a need to balance a high risk with a low probability of occurrence (that is children abused or groomed for abuse in a chat room) against a low risk with high occurrence (closing the chat rooms). **The rights of the majority against the risks of a minority.** There is no clear-cut right or wrong answer from looking at justice, equality and respect at present.